\documentclass[preprint,12pt]{elsarticle}
\usepackage[usenames,dvipsnames]{color}

\usepackage{graphicx}
\usepackage{amsmath}
\usepackage{amssymb}
\usepackage[version=4]{mhchem}
\usepackage{courier}
\usepackage{listings}
\usepackage{hyperref}
\usepackage[normalem]{ulem}

\newcommand{\cfont}[1]{{\fontfamily{pcr}\selectfont
#1}}

\definecolor{codegreen}{rgb}{0,0.6,0}
\definecolor{codegray}{rgb}{0.5,0.5,0.5}
\definecolor{codepurple}{rgb}{0.58,0,0.82}
\definecolor{backcolour}{rgb}{0.95,0.95,0.92}

\lstdefinestyle{mystyle}{
    backgroundcolor=\color{backcolour},   
    commentstyle=\color{codegreen},
    keywordstyle=\color{magenta},
    numberstyle=\tiny\color{codegray},
    stringstyle=\color{codepurple},
    basicstyle=\ttfamily\footnotesize,
    breakatwhitespace=false,         
    breaklines=true,                 
    captionpos=b,                    
    keepspaces=true,                 
    numbers=left,                    
    numbersep=5pt,                  
    showspaces=false,                
    showstringspaces=false,
    showtabs=false,                  
    tabsize=2
}

\newcounter{bla}

\journal{Computer Physics Communications}
\begin{document}

\begin{frontmatter}

\title{\cfont{Asparagus}: A Toolkit for Autonomous,
  User-Guided Construction of Machine-Learned Potential
  Energy Surfaces}

\author[a]{Kai T\"opfer\fnref{equal}\corref{author1}}
\author[a,b]{Luis Itza Vazquez-Salazar\fnref{equal}}
\author[a]{Markus Meuwly\corref{author2}}

\fntext[equal]{This authors contributed equally} 
\cortext[author1]
       {Corresponding author.\\\textit{E-mail address:}
         kai.toepfer@unibas.ch} \cortext[author2] {Corresponding
         author.\\\textit{E-mail address:} m.meuwly@unibas.ch}
       \address[a]{Department of Chemistry, University of Basel,
         Klingelbergstrasse 80, CH-4056 Basel, Switzerland.}
       \address[b]{Present Address: Institute of Theoretical Physics,
         Heidelberg University, Heidelberg, Germany.}
         
\begin{abstract}
With the establishment of machine learning (ML) techniques in the
scientific community, the construction of ML potential energy surfaces
(ML-PES) has become a standard process in physics and chemistry. So
far, improvements in the construction of ML-PES models have been
conducted independently, creating an initial hurdle for new users to
overcome and complicating the reproducibility of results.  Aiming to
reduce the bar for the extensive use of ML-PES, we introduce
\cfont{Asparagus}, a software package encompassing the different parts
into one coherent implementation that allows an autonomous,
user-guided construction of ML-PES models.  \cfont{Asparagus} combines
capabilities of initial data sampling with interfaces to {\it ab
  initio} calculation programs, ML model training, as well as model
evaluation and its application within other codes such as ASE or
CHARMM.  The functionalities of the code are illustrated in different
examples, including the dynamics of small molecules, the
representation of reactive potentials in organometallic compounds, and
atom diffusion on periodic surface structures.  The modular framework
of \cfont{Asparagus} is designed to allow simple implementations of
further ML-related methods and models to provide constant
user-friendly access to state-of-the-art ML techniques.
\end{abstract}


\end{frontmatter}

\noindent
{\bf PROGRAM SUMMARY}

\begin{small}
\noindent
    {\em Program Title:} \cfont{Asparagus}   \\
    
    {\em CPC Library link to program files:} (to be added by Technical Editor) \\
    
    {\em Developer's repository link:} \url{https://github.com/MMunibas/Asparagus} \\
    
    {\em Code Ocean capsule:} (to be added by Technical Editor)\\
    
    {\em Licensing provisions:} MIT  \\
    
    {\em Programming language:} Python                            \\
    
    {\em Supplementary material:} Access to Documentation at
    \url{https://asparagus-bundle.readthedocs.io}\\
    
    {\em Nature of the problem(approx. 50-250 words):}\\ Constructing
    machine-learning (ML) based potential energy surfaces (PESs) for
    atomistic simulations is a multi-step process that requires a
    broad knowledge in quantum chemistry, nuclear dynamics and
    programming.  So far, efforts mainly focused on developing and
    improving ML model architectures. However, there was less effort
    spent on providing tools for {\it consistent and reproducible
      workflows} that support the construction of ML-PES for a variety
    of chemical systems for the broader science community.\\

    {\em Solution method(approx. 50-250 words):}\\ \cfont{Asparagus}
    is a program package written in Python that provides a streamlined
    and extensible workflow with a user-friendly command structure to
    support the construction of ML-PESs. This is achieved by bundling
    and linking data generation and sampling techniques, data
    management, model training, testing and evaluation tools into one
    modular, comprehensive workflow including interfaces to other
    simulation packages for the application of the ML-PESs. By
    lowering the entrance barriers especially for new users,
    \cfont{Asparagus} supports the generation and adjustment of
    ML-PESs that allow an increased focus on the physico-chemical
    evaluation of the chemical system or application in molecular
    dynamics simulations.\\

    {\em Additional comments including restrictions and unusual
      features (approx. 50-250 words):}\\ \cfont{Asparagus} is a
    modular package written in Python providing an underlying
    structure for further extensions and maintenance. Currently,
    methods based on message-passing neural network (NN) models using
    the PyTorch Python package are available. Additions of new models
    and interfaces to already implemented modules are possible. The NN
    architecture and hyperparameters are stored in a global
    configuration module and as a \cfont{json} file for
    documentation. Except for essential input information, default
    input parameters are used if not specifically defined otherwise,
    which allows a quick setup for the construction of a ML-PES but
    also the fine-tuning for specific needs.\\

\end{small}
\setcounter{footnote}{0}

\section{Introduction}
Potential energy surfaces (PESs) for atomic systems are crucial for
investigating the structural and dynamical physico-chemical properties
of chemical systems at an atomistic level. Prerequisites for accurate
simulations are high-quality and validated representations of the
inter- and intramolecular interactions involved. Techniques for
constructing such machine-learned potential energy surfaces (ML-PES) -
also known as machine-learning potentials
\footnote{Although in the literature it is common to find both names,
the present work uses ML-PES instead of MLP to avoid confusion with
``multi-layer perceptron''.} - have gained traction over the past
decade. Representative approaches are based on permutationally
invariant polynomials (PIPs),\cite{houston2023pespip,bowman.irpc:2009}
neural network (NN) techniques as used in SchNet\cite{schnet:2018},
PhysNet,\cite{MM.physnet:2019} or DeepPot-SE,\cite{zhang:2018}
kernel-based methods, including (symmetrized) gradient-domain machine
learning ((s)GDML),\cite{chmiela:2017,sauceda:2019} reproducing kernel
Hilbert
spaces,\cite{rabitz:1996,hollebeek.annrevphychem.1999.rkhs,MM.rkhs:2017}
FCHL\cite{fchl:2020} or Gaussian process
regression\cite{bartok:2010,krems:2016}. The current state-of-the-art
of the field was also reviewed
continuously\cite{unke:2021,manzhos:2020,MM.cr:2021,deringer:2021}.\\

\noindent
Despite the progress that has been made, constructing and validating
ML-PESs suitable for MD or MC simulations can still be a
time-consuming and challenging task, particularly if globally robust
surfaces are sought. One particularly challenging application is the
study of chemical
reactions.\cite{unke2021machine,MM.cr:2021,qu:2021,MM.tl:2022} The
process of breaking and forming chemical bonds increases the
accessible configurational space that needs to be covered
dramatically. On the other hand, this is one of the applications where
ML-PESs ``shine'' as conventional parametrized PESs or more empirical
energy functions force fields are either not sufficiently accurate or
do not include the possibility of describing chemical reactions,
although exceptions
exist.\cite{warshel:1980,MM.msarmd:2014,vanduin:2001}\\

\noindent
One specific advantage of ML-based techniques is that the inference
time of such models is independent of the level of theory at which the
reference data - usually from electronic structure calculations - was
obtained. Once the reference data are available, the ML-PES is trained
and successfully validated, atomistic simulations can be carried out
significantly more efficiently and close to the corresponding level of
theory, which is particularly relevant for high-level reference
methods such as CCSD(T), CASPT2 or MRCI.\cite{jensen2017introduction}
Concerning inference times, NN-based representations are independent
on the size of the training set, whereas kernel-based methods scale
$\mathcal{O}(N_{\rm train}^{3})$ with training data set
size.\cite{pinheiro2023kernel} On the other hand, NN-based approaches
are in general more ``data hungry'' compared with kernel-based methods
in particular, if global and reactive PESs are sought.\\

\noindent
Program suites including TorchANI,\cite{gao2020torchani}
TorchMD,\cite{doerr2021torchmd} SchNetPack,\cite{schutt2023schnetpack}
FeNNol,\cite{ple2024fennol} MLAtom3,\cite{dral2024mlatom} and
DeePMD-kit\cite{wang2018deepmd,zeng2023deepmd} were introduced, which
allow training and using machine-learned PES models to run MD
simulations with in-built methods or provide interfaces to other
modules or programs, such as atomic simulation environment
(ASE)\cite{larsen2017atomic} or LAMMPS.\cite{LAMMPS} These programs,
however, require at least an initial set of reference training
data. In addition, and more relevant to the present work, a new
program suite ArcaNN\cite{david2024arcann} has been recently
introduced with the capability of generating the initial training sets
from \textit{ab-initio} MD simulations while using DeePMD-kit to
handle the ML-PES.\\

\noindent
The present work introduces the \cfont{Asparagus} suite that provides
Python-based utilities for automating the computational workflow to
generate high-dimensional ML-PESs from reference data acquisition,
model training and validation up to applications for property
predictions. \cfont{Asparagus} provides interfaces to Python packages
including ASE\cite{larsen2017atomic} and molecular simulation programs
such as CHARMM.\cite{pycharmm:2023,MM.pycharmm:2023} The present work
provides an overview of implemented functionalities and methods
together with illustrative application examples including input code
for reproduction.  \\

\section{Program Overview}
\label{sec:overview}
\cfont{Asparagus} is a workflow for constructing ML-PESs for given
molecular systems. Following the analogy of how an asparagus plant
grows, it is noticed that the different steps for constructing an
ML-PES are developed independently following a modular
fashion. \cfont{Asparagus} is completely written in Python and builds
on state-of-the-art tools and the most recent versions of
PyTorch\cite{paszke2019pytorch} and ASE\cite{larsen2017atomic}.\\

\noindent
The modular structure of \cfont{Asparagus} enables additions to cover
and include new developments and tools in each module without the need
to modify other modules or the parameter pipeline between them. The
input parameters for constructing a ML-PES are stored in a global
configuration module and written to a \cfont{json} file, which is
continuously updated for parameter documentation. The aim is to allow
reproducibility of the workflow or recovering the latest state of the
model.\\

\noindent
Constructing a ML-PES can be divided into several fundamental steps,
see Figure
\ref{fig:fig1}).\cite{behler2021four,unke2021machine,MM.rev:2023} The
strategies implemented and available in \cfont{Asparagus} are
described in the following.

\begin{enumerate}
\item \textbf{Sampling} Reference structures for an initial ML-PES
  first need to be generated. This can be accomplished with the
  various sampling methods implemented in \cfont{Asparagus}, or such
  samples can be imported from a pre-existing reference data source.
  \cfont{Asparagus} also supports the reference property computation
  of the sample structures with methods at a user-defined level of
  theory. In particular, the interface between \cfont{Asparagus} and
  ASE supports calculators for quantum chemical programs such as
  ORCA\cite{orca:2020} or Gaussian\cite{gaussian16} but can also be
  extended to other such programs. It also provides reference
  calculation schemes including empirical energy functions, density
  functional theory (DFT) and higher level {\it ab initio} methods
  based on preparing and running customizable template files to
  provide flexibility. The molecular structures and its reference
  properties, including energy, atomic forces, atom charges, and
  molecular dipoles, are stored in a \cfont{Asparagus} style database
  file in different formats, including SQL (default), HDF5 or Numpy
  npz.  \\

\item \textbf{Training} The database format implemented in
  \cfont{Asparagus} provides the information required for training an
  ML-PES. The reference data are split into a training, validation,
  and testing data subsets. The loss function needs to be defined, and
  a PyTorch optimizer is initialized, either from the default settings
  or through user-specific input. Currently, \cfont{Asparagus} is
  linked to the PhysNet\cite{unke2019physnet} and
  PaiNN\cite{schuett2021painn} NN architectures. The modular approach
  of \cfont{Asparagus}, however, allows for the straightforward
  addition of further established ML architectures such as
  SchNet\cite{schutt2018schnet}, Nequip\cite{nequip:2022}, or
  MACE\cite{batatia2022mace}.  Within the capabilities of the PyTorch
  module, the training procedure can be executed on the CPU or
  GPU. During training, the best model parameter set is stored
  according to the lowest loss value for the property prediction of
  the validation dataset.\\

\item \textbf{Testing} \cfont{Asparagus} provides functions to
  evaluate the accuracy of property predictions (see {\bf Sampling}
  above) for the complete reference dataset or its subsets. This
  module returns statistical measures such as mean absolute error
  (MAE) and root-mean squared error (RMSE) for each reference
  property. It is also possible to generate various correlation plots
  (reference vs. predicted property, reference vs. prediction error)
  or prediction error histograms. During model training, if a new best
  model parameter set is found at the model validation, the mentioned
  test functions are executed for the test data subset by default.\\

\item \textbf{Characterization} \cfont{Asparagus} includes native
  tools to determine important characteristics of a ML-PES. These
  tools allow searching for a minimum energy path or minimum dynamic
  path\cite{unke2019sampling} along the PES between two reactant
  states. A diffusion Monte-Carlo (DMC) method is implemented for the
  search of regions in the PES undersampled within the reference
  dataset.\cite{kosztin1996introduction,li2021diffusion,conte2020full,kaser2022transfer}.
  The ML-PES is available as an ASE calculator and can be used through
  ASE to determine, e.g. harmonic frequencies. This function can be
  used to further validate the accuracy and stability of the ML-PES or
  to identify regions in configurational space which require
  additional samples. In the latter case, these regions can be
  additionally sampled; the data is added to the reference dataset and
  used to refine the ML-PES.\cite{MM.pycharmm:2023}\\

\item \textbf{Interfaces} As already mentioned, the ML-PES can be
  loaded as an ASE calculator object and used for ASE functionalities
  such as MD simulations. Alternatively, MD simulations can be carried
  out through an interface between \cfont{Asparagus} and the CHARMM
  suite of codes through the MLpot module of the pyCHARMM API (see
  section \ref{sec:examples}).\cite{Brooks.charmm:2009,pycharmm:2023}
  This enables MD simulations using a) the ML-PES for the energy and
  force evaluation of a system or b) ML/MM simulations using
  mechanical embedding of the ML-PES and the CGenFF force
  field\cite{mackerell2010CgenFF} in CHARMM, e.g., for simulation of a
  molecule (ML) in solution (MM).\cite{MM.fad:2022} \\
\end{enumerate}

\section{Program Features}
\label{sec:features}
\begin{figure}[h!]
\centering
\includegraphics[width=\textwidth]{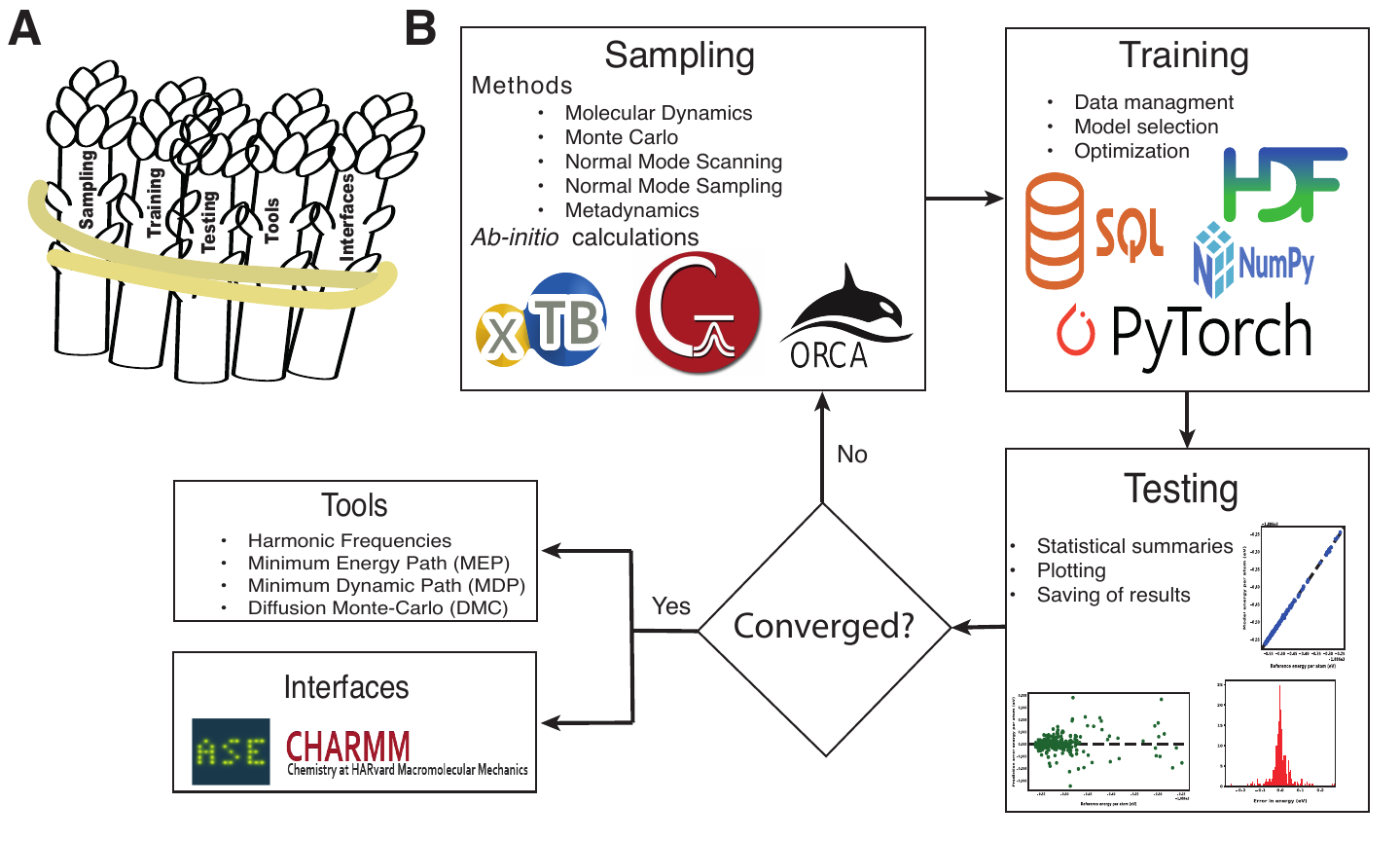}
\caption{Scheme representation of the \cfont{Asparagus} classes (panel
  A) and a workflow chart (panel B) that represents class details and
  procedures for the construction of a ML-PES in \cfont{Asparagus}.}
\label{fig:fig1}
\end{figure}

As summarized in Section \ref{sec:overview}, \cfont{Asparagus}
functions can be divided into five main classes: sampling, training,
testing, characterization and interfaces. A schematic overview of the
functions and their interdependencies is shown in Figure
\ref{fig:fig1}. In the following, the capabilities of each class are
described in more detail. Further information can be found in online
documentation (\url{https://asparagus-bundle.readthedocs.io/}).

\subsection{Sampling}
Sampling the configurational space is a key step for constructing an
ML-PES since the task is to deduce a functional form of the PES purely
from data.\cite{behler2021four,MM.rev:2023} Nevertheless, an
exhaustive sampling of the configuration space is usually unfeasible
because of the exponential scaling of the size of configurational
space, in particular for applications to chemical
reactions. Additionally, the computational cost of quantum chemical
calculations also scale exponentially with the number of electrons,
which becomes computationally significant when high-level methods such
as CCSD(T) or multi-reference methods including CASPT2 or MRCI, are
applied.\\

\noindent
There are multiple strategies to perform the initial sampling of an
atomic system. In \cfont{Asparagus}, some of the most commonly used
methods are implemented, including molecular dynamics (MD),
Monte-Carlo (MC), normal mode sampling (NMSamp), and metadynamics
(Meta). Additionally, an alternative to NMSamp is available, which is
referred to as normal mode scanning (NMScan). The implementation of
the sampling methods makes extensive use of ASE. Therefore, other
initialized ASE optimizer or propagator objects, such as for nudged
elastic band (NEB)\cite{henkelman2002methods}, minima
hooping\cite{krummenacher2024performing}, or basin
hopping\cite{wales1997global} can be passed to \cfont{Asparagus} to
run and store the reference data.\\

\noindent
By default, \cfont{Asparagus} uses the ASE interface of the
xTB\cite{gfn2xtb2019} calculator to compute initial reference
properties. This tight binding density functional method GFN2-xTB is
fast and convenient but will not provide accurate reference properties
for training a ML-PES. Still, it is useful for test purposes of the
command pipeline. For high-level results, ASE calculators connecting
to \textit{ab-initio} codes such as ORCA, MOLPRO,\cite{MOLPRO_brief}
or Gaussian should be used. The following provides background on the
sampling methods implemented in \cfont{Asparagus}.\\

\paragraph{Molecular Dynamics and Monte-Carlo}
Molecular Dynamics (MD) or Monte-Carlo (MC) simulations are the most
common methods to generate initial databases to construct a
ML-PES. For MD simulations, the Newtonian equations of motion are
propagated in time to explore the configuration space of a chemical
system. The given temperature $T$ will not only define the average
kinetic energy of the system but also determine the magnitude of
deformations of the atomic systems conformations. This also affects
the possibility of overcoming reaction barriers that yield sampling of
a larger configuration space. Usually, the sampling of the chemical
system should be carried out at a higher temperature than that
envisaged for the production simulations. This ensures that the ML-PES
does not enter the extrapolation regime during simulations for which
especially NN-based PESs cannot guarantee accurate property
predictions.\cite{MM.rev:2023,unke2021machine} \\

\noindent
Although MD is commonly used for the initial sampling of PESs for ML
models, it has some disadvantages, such as the correlation between the
generated structures and a negligible probability of sampling rare
events (i.e. reactive systems).\cite{unke2021machine} Therefore, MD
should be used in simulations close to equilibrium that do not involve
rare events.\cite{unke2021machine} In \cfont{Asparagus}, MD sampling
is implemented using Langevin dynamics, where the temperature of the
system is kept constant by adding a fluctuating force and a friction
term is used to emulate the presence of a heat bath.  \\

\noindent
On the other hand, MC sampling is a method where the configuration
space $\boldsymbol{x}$ is explored by random atom
displacements.\cite{jensen2017introduction} The MC method generates
random single-atom displacements $\boldsymbol{x'}$ with respect to a
uniform distribution. The new position is accepted if the energy
difference of the system $\Delta E = E(\boldsymbol{x'}) -
E(\boldsymbol{x})$ in the form of $a = \exp(-\Delta E /
k_\mathrm{b}T)$ is smaller than a random value $c = [0, 1]$ of uniform
distribution. The acceptance criteria $a$ can be modulated by changing
the sampling temperature. This implementation is known as the
Metropolis-Hastings (MH) algorithm.\cite{chib:1995} The MH algorithm
is commonly used in molecular simulations and implemented in
\cfont{Asparagus}. In general, MC has the advantage over MD in that MC
does not require any forces.\cite{metropolis1953}\\

\paragraph{Normal Mode Sampling}
Normal mode sampling (NMSamp) is an alternative to MD-based sampling
and allows targeted characterization of relevant regions of a
PES.\cite{smith2017ani,MM.rev:2023} Using the vibrational normal mode
vectors $\boldsymbol{Q} = {\boldsymbol{q}_i}$ obtained from harmonic
analysis of a molecule in an equilibrium conformations
$\boldsymbol{x_\mathrm{eq}}$, NMSamp generates new sample
conformations $\boldsymbol{x_n}$ in a random fashion by applying all
$N_\nu$ normal mode vectors, each scaled by a factor $f_i(c_i, k_i,
T)$
\begin{equation}
    \boldsymbol{x}_n = \boldsymbol{x}_\mathrm{eq} + \sum_i^{N_\nu}
    f_i(c_i, k_i, T) \boldsymbol{q}_i
\end{equation}
applied to the equilibrium conformation. The displacement scaling
factor
\begin{equation}
    f_i = \pm \sqrt{\frac{3 c_i k_\mathrm{b} T}{k_i}}
\end{equation}
depends on the random number $c_i \in [0, 1]$, the force constant
$k_i$ of the respective normal mode $i$, temperature $T$, and
$k_\mathrm{b}$ is the Boltzmann constant. The sign of $f_i$ is
determined randomly from a Bernoulli distribution with $P=0.5$. The
procedure is repeated until the desired number of samples has been
generated.\\

\noindent
NMSamp generates uncorrelated conformations in an efficient
manner. However, the sampling is based on the harmonic approximation
around the minimum energy structure. NMSamp can also be combined with
other techniques, such as NEB, to sample regions along a specific path
in the PES, which becomes particularly convenient for reactive
systems.\cite{brezina2023reducing}\\

\paragraph{Normal Mode Scanning}
Normal mode scanning (NMScan) is a sampling algorithm that generates
atom displacements along scaled vibrational normal mode vectors
$\boldsymbol{Q_\nu} = \{ \boldsymbol{q}_i \}$ on the initial
conformation $\boldsymbol{x}_\mathrm{init}$ for which the harmonic
frequencies $\nu$ and normal modes have been computed. The algorithm
iterates over a specified combination of normal modes and applies a
normal mode vector (or combination of normal mode vectors) scaled by
negative and positive multiples $n$ of a frequency-dependent scaling
factor $f_i$. The scaling factor $f_i$ is determined by a user-defined
input $E_\mathrm{step}$.  Within the harmonic approximation, the
energy step size $E_\mathrm{step}$ is supposed to match the energy
difference $\Delta E_{i,n=\pm1}(f_i)$
\begin{equation}
    E_\mathrm{step} = \Delta E_{i,n=\pm1}(f_i) = |
    E(\boldsymbol{x}_\mathrm{init} \pm f_i
    \boldsymbol{q}_i) - E(\boldsymbol{x}_\mathrm{init}) |
\end{equation}
between the initial energy $E(\boldsymbol{x}_\mathrm{init})$ and the
energy when the respectively scaled normal mode vector is applied once
$E(\boldsymbol{x}_\mathrm{init} \pm f_i \boldsymbol{q}_i)$. Positive
and negative multiples of the scaled normal mode vector ($n f_i
\boldsymbol{q}_i$ with $n \subset \mathbb{Z}$) are applied until the
absolute energy difference $\Delta E_{\nu,n}$
\begin{equation}
    \Delta E_{i,n} = | E(\boldsymbol{x}_\mathrm{init} + n f_i
    \boldsymbol{q}_i) - E(\boldsymbol{x}_\mathrm{init}) |
\end{equation}
reaches a user-defined energy limit $E_\mathrm{limit}$ or the absolute
value of the multiplier $|n|$ reached a user-defined step limit
$n_\mathrm{max}$.  \\

\noindent
The normal mode scaling factor $f_i$ for each frequency $\nu_i$
depends on their respective force constant $k_i$
\begin{equation}
    k_i = 4\pi^2 (c\nu_i)^2/\mu_i
\end{equation}
where $c$ is the speed of light and $\mu_i$ is the reduced mass of the
respective normal mode
\begin{equation}
  \mu_i = \left[ \sum_j^{N_\mathrm{atoms}} \left( \boldsymbol{q}_i
    \cdot \boldsymbol{q}_i \right) / m_j \right]^{-1}
\end{equation}
with $m_j$ the atom mass of atom $j$. The normal modes are each
normalized by $\boldsymbol{q}_i = \boldsymbol{q'}_i / \left| \left|
\boldsymbol{q'}_i \right| \right|$ with $\boldsymbol{q'}_i$ as the
normal mode vector from the harmonic vibrational analysis.  According
to the harmonic approximation for the potential along the scaled
normal mode vector $f_i \boldsymbol{q}_i$ with
\begin{equation}
    E_i(f) = 0.5 k_i \cdot \left| \left| f \boldsymbol{q}_i \right|
    \right|^2 = 0.5 k_i \cdot f^2 ~~~,
\end{equation}
the respective scaling factor $f_i$ to yield $E_\mathrm{step}$ is
\begin{equation}
    f_i = \sqrt{\dfrac{2 E_\mathrm{step}}{k_i}}.
\end{equation}

\noindent
It is important to mention that the real initial energy difference
($n=\pm1$) will not yield the defined energy step value
$E_\mathrm{step}$ due to the anharmonicity of the PES and will not
change proportional to $\propto n^2$ for increasing $n$. Depending on
the atomic system and vibrational mode, the real energy steps can both
propagate less strongly (e.g. bond dissociation of a diatomic
molecule) or more strongly (e.g. bending modes of larger
molecules). It is, therefore, not possible to predict the exact number
of steps computed along the scan path until the energy limit is
reached.\\

\noindent
The initial conformation $\boldsymbol{x}_\mathrm{init}$ on which the
harmonic normal mode analysis is applied does not necessarily have to
be an equilibrium structure. Even for structures exhibiting one
(transition state) or multiple imaginary frequencies, the algorithm
handles the normal modes as usual until the energy difference $\Delta
E_{i,n}$, defined as absolute value reaches the energy limit or step
limit.  \\

\noindent
As mentioned, normal mode scanning can applied to a combination
$\boldsymbol{C}$ of a number multiple normal modes $N_C$, that yields
an expression for the energy difference
\begin{equation}
    \Delta E_{\boldsymbol{C},\boldsymbol{n}} = |
    E(\boldsymbol{x}_\mathrm{init} + \sum_{i,i\in
      \boldsymbol{C}}^{N_C} n_i f_i \boldsymbol{q}_i) -
    E(\boldsymbol{x}_\mathrm{init}) |
\end{equation}
with multipliers $\boldsymbol{n} = \{ n_i \}$.  However, iterating
over all possible combinations of normal modes lead to a large number
of new conformations, which can become excessive for larger
molecules. In practice, it may be advantageous to apply a scan over a
combination of a subset of normal modes with, e.g., frequencies in a
certain range of wave numbers.  \\

\paragraph{Metadynamics}
Metadynamics is a technique that allows the acceleration of sampling
rare events and the estimation of the free energies of complex atomic
systems.\cite{laio2002meta,bussi2020using} This is achieved by
introducing a history-dependent bias potential based on a number
$N_\mathrm{cv}$ of collective variables (CVs) denoted as
$\boldsymbol{S} = \{ s_i \}$. In general, a CV can be any of the
degrees of freedom of the
system.\cite{peters2016reaction,herr2018meta,pfaendtner2019metadynamics}
In \cfont{Asparagus} currently supported CVs are bond
distances, bond angles, dihedral angles or reactive coordinates
(e.g. difference between two bond distances) between atoms. The choice
of CVs crucially affects the effectiveness of metadynamics to yield
reliable free energy surfaces for rare events.\cite{bussi2020using}
For conformational sampling the selection of CVs is less critical as
their choice should just provide sufficient coverage of the structures
along the ``reaction''
path.\cite{unke2021machine,herr2018meta,yoo2021metasampl} \\

\noindent
In practice, Metadynamics simulations are MD (or MC) simulations on a
PES $V(\boldsymbol{x})$ that is perturbed by a ``bump'' potential
$V_\mathrm{bump}$ to gradually fill the basins of $V(\boldsymbol{x})$.
$V_\mathrm{bump}$ is is a sum of $N_\alpha$ Gaussians
\begin{equation}
    V_\mathrm{bump} = \sum_\alpha^{N_\alpha} G(\boldsymbol{S},
    \boldsymbol{S}_\alpha) = \sum_\alpha^{N_\alpha} \lambda \exp
    \left( \sum_i^{N_\mathrm{cv}} \left( s_i - s_{i,\alpha} \right)^2
    / \left( 2\sigma_i^2 \right) \right)
\end{equation}
each centered around CV coordinates $\boldsymbol{s}_\alpha$ with
Gaussian height $\lambda$ and width $\boldsymbol{\sigma} = \{ \sigma_i
\}$. The number of Gaussians $N_\alpha$ increases by one at each
user-defined interval of the simulation with a new set of
$\boldsymbol{S}_\alpha$ at the current frame.  \\

\noindent
Usually, the combination of Gaussian heights $\lambda$ and widths
$\boldsymbol{\sigma}$ is chosen small enough to keep the system in
close to thermodynamic equilibrium but sufficiently large to achieve
efficient sampling of the free energy surface.\cite{bussi2020using}
For the purpose of conformation sampling, keeping the thermodynamic
equilibrium is less relevant and large $\lambda$ and
$\boldsymbol{\sigma}$ values yield rapid sampling of the configuration
space. The consequently large increase in the simulation temperature
is countered by a high friction coefficient of the applied Langevin
thermostat. For that reason, the implemented sampling algorithm is
called meta-sampling rather than -dynamics.  \\

\subsection{Training and Testing}
The reference data used for training and testing of the ML-PES are
stored in the \cfont{Asparagus} database file.  The database class of
\cfont{Asparagus} is inspired by the database class implemented in
ASE, providing a minimalist version to store, update and supply
data. It is designed to allow data to be stored in different file
formats such as SQL (via sqlite3 python package), HDF5 (via h5py
python package), or \cfont{.npz} (compressed numpy format)
format. Reference data can be read from all \cfont{Asparagus} database
formats and formats such as ASE database files, ASE trajectory files
and Numpy npz files. \\

\noindent
The database entries contain general information on atom types,
positions and the system's total charge. Additionally, support for
periodic systems is provided by storing boundary conditions in each
Cartesian direction and cell parameters. \cfont{Asparagus} provides
the ``quality-of-life feature'' to internally handle unit conversion
between different reference datasets according to the defined property
units.  It will also handle the conversion between reference property
units and different model property units during the model training if
needed. By default, ASE units will be used, i.e. positions in \AA\/
and energies in eV.\\

\noindent
If not specifically defined, the default settings and hyperparameters
are used for the training procedure. By default, the reference data
are split into 80\% for the training, 10\% for validation, and 10\%
for testing. The loss function for parameter optimization is
\begin{equation}
\begin{aligned}
     \mathcal{L} = W_{E}\mathcal{L}_{E}(E_{\rm ref},E_{\rm pred}) \ +
     \ W_{F}\mathcal{L}_{F}(F_{{\rm ref}},F_{\rm pred}) \ +\\ W_{D}
     \mathcal{L}_{D}(D_{\rm ref},D_{\rm pred}) \ + \ W_{Q}
     \mathcal{L}_{Q}(Q_{\rm ref}, Q_{\rm pred})
\end{aligned}
\label{eq:loss_f}
\end{equation}
Here, the weights $W \in [E,F,D,Q]$ for the properties energy $E$,
forces $F$, molecular dipole $D$ and atomic charges $Q$ are defined as
$W_{E}=1$, $W_{F}=50$, $W_{D}=25$ and $W_{Q}=1$.  The smooth L1 loss
function $\mathcal{L}$ is defined as:
\begin{equation}
    \mathcal{L}(x_{\rm ref},x_{\rm pred}) = \left\{
    \begin{array}{ll}
         0.5 \cdot \dfrac{(x_{\rm ref}-x_{\rm pred})^{2}}{\beta} &
         {\rm If} \ |x_{\rm ref}-x_{\rm pred}|<\beta \\ |x_{\rm
           ref}-x_{\rm pred}| - 0.5 \cdot \beta & {\rm else}
    \end{array}
    \right.
\end{equation}
where $\beta=1$, although it can be adjusted by the user. Also, the
user can define other functional forms for the loss function. By
default, \cfont{Asparagus} uses the AMSGrad variant of the Adam
algorithm\cite{kingma2014adam,j.2018on} with a learning rate of 0.001,
a weight decay of $10^{-5}$, and an exponential learning rate
scheduler with a decay rate of 0.999. All optimization algorithms and
learning rate schedulers implemented in PyTorch are available. To avoid
overfitting, early stopping,\cite{prechelt2002early} exponential
moving average with a decay of 0.99, and a gradient clipping function
are implemented. The training is performed in epochs with a validation
step every $n$th epoch ($n=5$ by default). Checkpoint files of the
model parameter state are stored for the model at the last validation
step, and each time a new best-performing validation loss value is
reached.  \\

\noindent
After training, each best model state is evaluated on the test data.
\cfont{Asparagus} computes statistical quantities including MAE, RMSE
and the Pearson correlation coefficient ($1 - r^{2}$) for each
property of the loss function. The performance of the trained model is
graphically represented as $(x/y)-$correlation plots between reference
values $(x)$ and property prediction $(y)$ and reference values,
prediction error and histograms of prediction error.\\

\subsection{Tools}
Once an ML-PES has been obtained, standard properties that
characterize it can be determined. In \cfont{Asparagus}, suitable
tools have been implemented, which are briefly described in the
following.  It should be noted, however, that other quantities than
those discussed next may be of interest to the user, depending on the
project at hand.\\

\paragraph{Minimum Energy Path}
The MEP is defined as the lowest path energy connecting reactants and
products by passing through the transition state. The MEP is obtained
by following the negative gradient of the PES starting from the
transition state along the normal coordinate of the imaginary
frequency. Usually, this is done by integrating the path in small
steps ($\epsilon$) using the Euler method, which updates the positions
as:
\begin{equation}
    \boldsymbol{x}_{n+1} = \boldsymbol{x}_{n} - \epsilon\nabla
    V(\boldsymbol{x}_{n}) \ ; \ \boldsymbol{x}_{0} =
    \boldsymbol{x}_{\rm TS}
\end{equation}

\paragraph{Minimum Dynamic Path}
Complementary to the minimum energy path is the minimum dynamic path
(MDP),\cite{unke2019sampling} which provides information about the
lowest path between reactants and products in phase space. In this
case, Newton's equation of motion is integrated over the normal
coordinate of the imaginary frequency of the TS for small time steps
($\epsilon$) using the velocity Verlet scheme\cite{verlet1967}. This
formulation keeps information about the previous gradients in the
velocities. Then, the positions and velocities are obtained as:
\begin{align}
        \boldsymbol{x}_{n+1} = \boldsymbol{x}_{n} + \epsilon
        \boldsymbol{v}_{n} - \frac{1}{2m}\nabla V(\boldsymbol{x}_{n})
        \ ; \ \boldsymbol{x}_{0} = \boldsymbol{x}_{TS}
        \\ \boldsymbol{v}_{n+1} = \boldsymbol{v}_{n} -
        \frac{\epsilon}{2m}(\nabla V(\boldsymbol{x}_{n}) + \nabla
        V(\boldsymbol{x}_{n+1})) \ ; \ \boldsymbol{v}_{0}=0
\end{align}

\paragraph{Diffusion Monte Carlo}
DMC is based on the similarity between the diffusion equation with a
sink term and the imaginary-time Schr\"odinger equation (replace $t
\rightarrow -i \tau$) with an energy shift term.\cite{li2021diffusion}
Then, random-walk simulations can be used to solve it and to obtain
the quantum mechanical zero-point energy (ZPE) and nuclear
ground-state wavefunction of a
molecule\cite{anderson1975random,quack1991potential,kosztin1996introduction}. During
a DMC simulation, the atoms are randomly displaced, allowing an
efficient exploration of conformational space. Therefore, the DMC
method can be used to detect holes (i.e. regions on a PES that have
large negative energies with respect to the global minimum) in
ML-PESs.\cite{conte2020full} \cfont{Asparagus} uses DMC for this
purpose.\\

\noindent
The method is formulated as follows. For a system of interest, a set
of walkers - their ensemble is a proxy for the nuclear wavefunction -
is initialized at $x_{0}$. The walkers are then randomly displaced at
each time step $\tau$ according to
\begin{align}\label{eq:dmc_disp}
    \boldsymbol{x}_{\tau + \Delta\tau} = \boldsymbol{x}_\tau +
    \sqrt{\frac{\hbar \Delta\tau}{m}r}
\end{align}
where $\boldsymbol{x}_\tau$ corresponds to coordinates at time step
$\tau$, $\Delta\tau$ is the time step of the random-walk simulation,
$m$ is the atomic mass, and $r$ is a random number drawn from a
Gaussian distribution, $\mathcal{N}(0,1)$. The walkers obtained from
Equation \ref{eq:dmc_disp} are then used to compute the potential
energy ($E_{i}$) of each walker $i$. In the next step, the potential
energies of the walkers are compared with a reference energy $E_{r}$,
to determine whether a walker stays alive, gives birth to a new walker
or is killed. The probabilities of birth or death of walker are given
by
\begin{align}
    P_{\rm death} = 1 - e^{-(E_i -E_r)\Delta\tau} \quad (E_i > E_r)\\
    P_{\rm birth} = e^{-(E_i -E_r)\Delta\tau} - 1 \quad (E_i < E_r)
\end{align}
After the probabilities are obtained, the walkers that do not pass the
threshold are removed, and new walkers are born. As a consequence of
the dead-birth process, the number of alive walkers fluctuates. Next,
$E_r$ is adjusted according to
\begin{align}\label{eq:update_dmc}
    E_r(\tau) = \left<V(\tau)\right> - \alpha \frac{N(\tau) -
      N(0)}{N(0)}.
\end{align}
The averaged potential energy of the alive walkers is given by
$\left<V(\tau)\right>$, $\alpha$ is a constant/parameter that governs
the fluctuation in the number of walkers and the reference energy, and
$N(\tau)$ and $N(0)$ are the number of alive walkers at time step
$\tau$ and 0, respectively.

\section{Examples of Use}
In the following, representative applications are described. The first
example is discussed in more detail together with sample code, whereas
the second and third examples are more illustrative of the
capabilities of \cfont{Asparagus}.

\label{sec:examples}
\subsection{Conformational Sampling in the Gas Phase and in Solution: Ammonia}
\label{ex1}
Conformations for ammonia (NH$_3$) were sampled by different sampling
methods implemented in \cfont{Asparagus} and used to train a ML-PES
using PhysNet to show the capabilities and limitations of the sampling
methods and their impact on a trained ML-PES. The performance of the
models was evaluated by the RMSE between model and reference
properties, bond elongation potentials, vibrational harmonic
frequencies and simulation results of single ammonia in water using a
QM(ML-PES)/MM approach with mechanical embedding.  \\

\noindent
ML-PESs were trained using \cfont{Asparagus} and different reference
data for a single ammonia molecule sampled by (A) MD simulation at
500\,K (Listing \ref{code:md_nh3}), (B) Metadynamics sampling (Listing
\ref{code:meta_nh3}) at 500 K with each N-H bond assigned as CV, and
(C) normal mode scanning along single (Listing \ref{code:nmscan_nh3})
and permutations of two normal mode vectors. By providing an initial
guess of the ammonia structure (e.g. via a \emph{.xyz} file), all
sampling methods were initialized and started by a single Python
command, respectively.\\

\begin{lstlisting}[
language=Python, 
caption=Input for MD sampling of NH$_3$ at 500\,K producing 
1000 reference samples,
label={code:md_nh3}]
from asparagus.sample import MDSampler
sampler = MDSampler(
    sample_systems='nh3_c3v.xyz',
    sample_calculator='ORCA',
    sample_calculator_args = {
        'orcasimpleinput': 'RI PBE D3BJ def2-SVP def2/J'},
    md_temperature=500, # K
    md_time_step=1.0, # fs
    md_simulation_time=10_000.0, # fs
    md_save_interval=10,
    )
sampler.run()
\end{lstlisting}

\begin{lstlisting}[
language=Python, 
caption=Input for metadynamics sampling of NH$_3$ with the 3 N-H bonds
as collective variables producing 1000 reference samples,
label={code:meta_nh3}]
from asparagus.sample import MetaSampler
sampler = MetaSampler(
    sample_systems='nh3_c3v.xyz',
    sample_calculator='ORCA',
    sample_calculator_args = {
        'orcasimpleinput': 'RI PBE D3BJ def2-SVP def2/J'},
    meta_cv=[[0, 1], [0, 2], [0, 3]], # N(0)-H(1,2,3)
    meta_gaussian_height=0.05, # eV
    meta_gaussian_widths=0.1, # Angstrom
    meta_gaussian_interval=10,
    meta_temperature=500, # K
    meta_time_step=1.0, # fs
    meta_simulation_time=10_000.0, # fs
    meta_save_interval=10,
    )
sampler.run()
\end{lstlisting}

\begin{lstlisting}[
language=Python, 
caption=Input for normal mode scanning of NH$_3$ along
single and a combination of two normal modes that
produces 1595 reference samples,
label={code:nmscan_nh3}]
from asparagus.sample import NormalModeScanner
sampler = NormalModeScanner(
    sample_systems='nh3_c3v.xyz',
    sample_calculator='ORCA',
    sample_calculator_args = {
        'orcasimpleinput': 'RI PBE D3BJ def2-SVP def2/J'},
    nms_harmonic_energy_step=0.05, # eV
    nms_energy_limits=1.00, # eV
    nms_number_of_coupling=2,
    )
sampler.run(nms_frequency_range=[('>', 100)]) # modes > 100cm-1
\end{lstlisting}

The sampling methods generated 1000 (method A), 1000 (B) and 1595 (C)
structures for which reference data (energies, forces, dipole moments)
was generated at the PBE-D3/def2-SVP level using ORCA.\cite{orca:2020}
As indicated above, the final number of samples generated by normal
mode scanning can not be predicted {\it a priori}. Hence, the
difference between the number of samples between methods A/B and C.\\

\begin{lstlisting}[
language=Python, 
caption=Input for PhysNet model training and final evaluation
on all datasubsets,
label={code:train_nh3}]
from asparagus import Asparagus
model = Asparagus(
    config='config.json', # File path to store model parameters
    data_file='data.db', # Respective reference database
    model_type='physnet', # Default ML-PES
    model_properties=['energy', 'forces', 'dipole'],
    )
model.train(trainer_max_epochs=1_000)
model.test(test_datasets='all')
\end{lstlisting}

\begin{lstlisting}[
language=Python, 
caption=Input for PhysNet model evaluation by reading
parameters from the configuration file and loading the
ASE calculator interface,
label={code:calc_nh3}]
from asparagus import Asparagus
from ase import io
model = Asparagus(config='config.json')
calc = model.get_ase_calculator()
ammonia = io.read('nh3_c3v.xyz', format='xyz')
ammonia.calc = calc
ammonia.get_potential_energy()
# ... model potential evaluation
\end{lstlisting}

\begin{figure}[h!]
\centering
\includegraphics[width=0.9\textwidth]{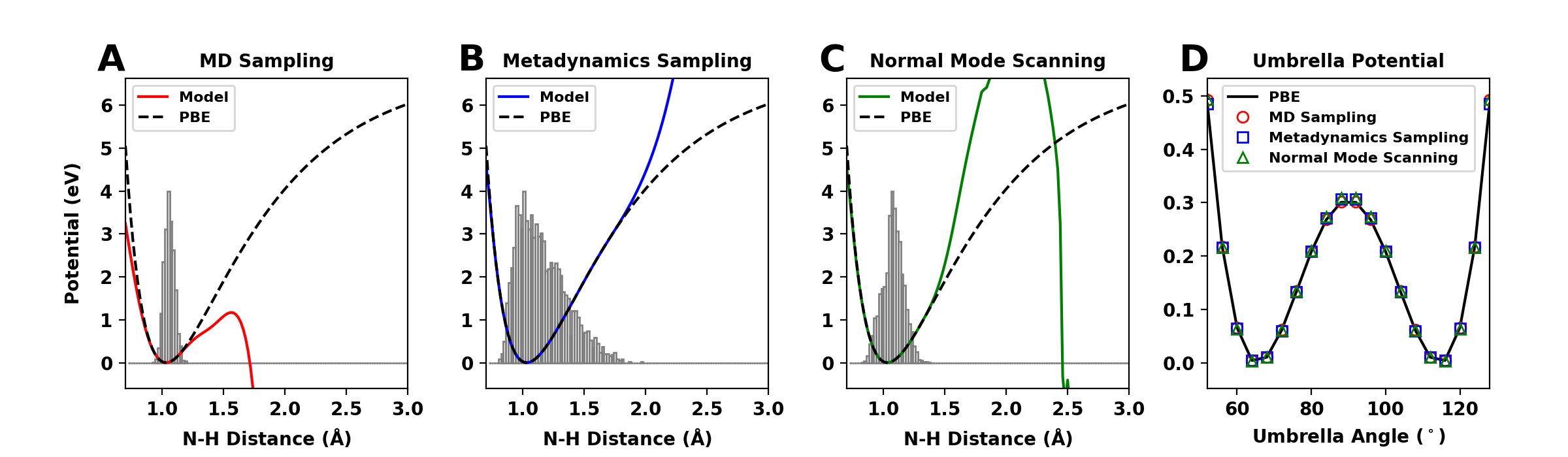}
\caption{Bond potential for a single N-H bond elongation in ammonia
  predicted by the PBE reference method (dashed black line) and
  PhysNet model potentials trained on reference data from (A) MD
  simulation, (B) metadynamics and (C) normal mode scanning sampling
  methods. The hollow grey bars indicate N-H bond distance
  distribution in the respective data sets.  Panel D shows the
  potential curve along the umbrella motion in ammonia predicted by
  the PBE reference method (dashed black line) and the PhysNet model
  potentials trained on reference data from MD simulation (red line),
  metadynamics (blue dash-dotted line) and normal mode scanning
  sampling methods (green dotted line).}
\label{fig:nh_bond}
\end{figure}

\noindent
The PhysNet ML-PESs were trained on energies, forces and molecular
dipole moments for at most 1000 epochs (sufficient for convergence)
using the input code in Listing \ref{code:train_nh3} for each of the
sampled reference data sets A to C.  Average RMSEs per conformation
and atoms for the energies (forces) on the test sets were (A)
$4.1$\,meV ($45.4$\,meV/\AA), (B) $7.4$\,meV ($59.3$\,meV/\AA) and (C)
$7.3$\,meV ($48.5$\,meV/\AA).  As judged from the RMSE values, the PES
trained on (B, metadynamics) performs worst for both energy and
forces. On the other hand, this data set covered the widest range of
potential energies compared with A or C, respectively, see Figure
\ref{fig:nh_bond}B. Further evaluations were performed by using ASE
and the ASE calculator interface of the model potential; see Listing
\ref{code:calc_nh3}.\\

\noindent
The advantage of sampling wider energy ranges can be seen in the
stretch potential of a single N-H bond of ammonia away from the
equilibrium conformation; see Figure \ref{fig:nh_bond}. Predictions
for the ML-PES trained on (B, metadynamics data) remain at least
qualitatively correct for N-H bond lengths up to $1.7$\,\AA\/ and
$3$\,eV above the minimum energy. For the curves in Figures
\ref{fig:nh_bond}A and \ref{fig:nh_bond}C the predictions start to
differ significantly from the reference potential at even smaller bond
elongation away from the equilibrium distance because larger distances
were not sufficiently covered in the reference data sets.\\

\noindent
For energy predictions along the umbrella motion in ammonia ($C_{3v}
\rightarrow D_{3h} \rightarrow C_{3v}$), see Figure
\ref{fig:nh_bond}D, the PhysNet model trained on MD data (A) predicts
closest to the reference energy difference between equilibrium and
transition state potential with an error of $0.15$\,meV. Barrier
heights for the ML-PESs trained on metadynamics sampling (B) and
normal mode scanning data (C) deviate by $5.04$\,meV and $6.84$\,meV,
respectively, which were also found to fluctuate more within a set of
independently trained models following the same workflow. The energy
RMSEs in a set of independently trained models on MD data remain
narrowly small. This may indicate an insufficient number of samples in
the metadynamics and NMS datasets with respect to the range of the
sampled configurational space or the potential energy range, which is
required for a well-trained and converged PhysNet model.\\

\noindent
Validation with respect to harmonic frequencies of the six vibrational
normal modes are considered next. The RMSE per normal mode between the
reference frequencies at the PBE level of theory and the model
frequencies are lowest ($16.0$\,cm$^{-1}$) for the model trained on
NMS data. ML-PESs trained on metadynamics sampling and MD data yield
an RMSE per normal mode of $31.6$\,cm$^{-1}$ and $53.8$\,cm$^{-1}$,
respectively. These differences indicate that further improvements can
be achieved by either adding additional samples around the global
minimum, more extensive training, or both. The model trained on
metadynamics data performs better for the higher N-H stretch mode
frequencies but worse for the lower bending mode frequencies compared
with the MD-data trained model.\\

\begin{lstlisting}[
language=Python, 
caption=Extraction of the PyCHARMM script to activate
the QM(ML)/MM potential representation for ammonia,
label={code:pycharmm}]
import pycharmm
from asparagus import Asparagus
from ase import io
# ... initialize simulation system and parameters
ml_model = Asparagus(config='config.json')
ammonia = io.read('nh3_c3v.xyz', format='xyz')
calc = pycharmm.MLpot(
    ml_model,
    # atomic numbers of ammonia: [7, 1, 1, 1]
    ammonia.get_atomic_numbers(),
    # segment name of ammonia: 'AMM1' 
    pycharmm.SelectAtoms(seg_id='AMM1'), 
    ml_charge=0, # total charge 0 (default)
    ml_fq=True) # use fluctuating model charges (default)
# ... continue with further simulation commands
\end{lstlisting}

\begin{figure}[h!]
\centering
\includegraphics[width=0.9\textwidth]{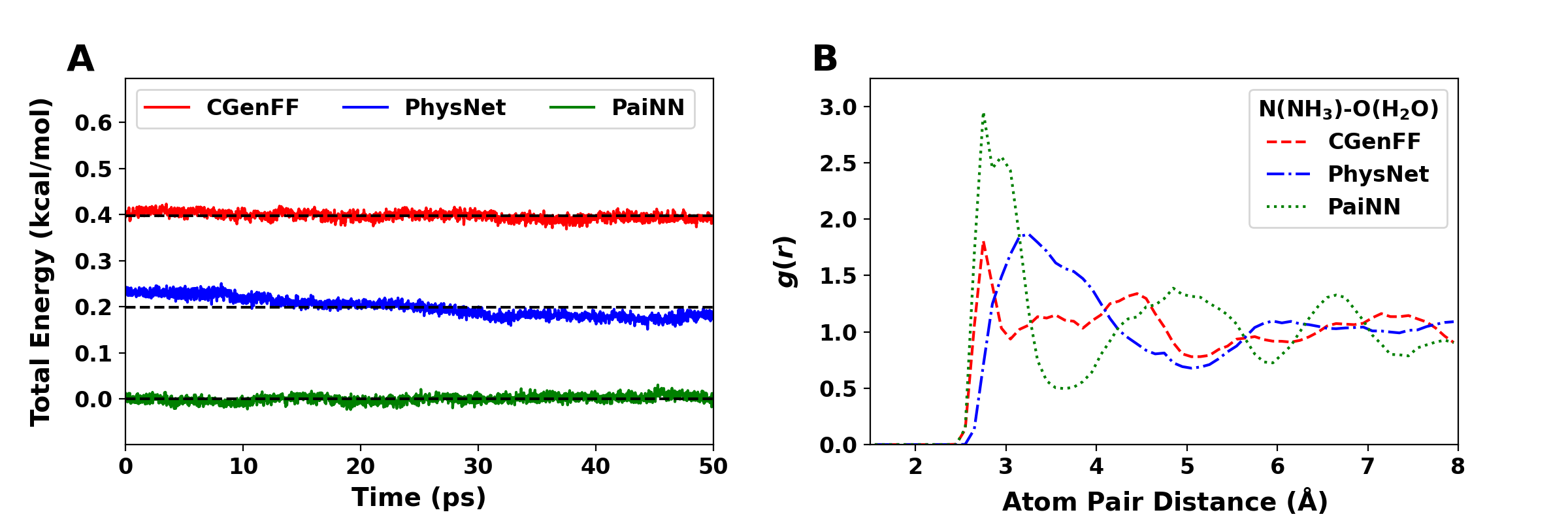}
\caption{Panel A: Total energy sequence of a $NVE$ simulation for
  ammonia in water using the classical force field CGenFF and the
  QM(ML)/MM approach with a trained PhysNet or PaiNN model of ammonia
  trained on metadynamics sampled data (see Listing
  \ref{code:meta_nh3}).  The energies are arbitrarily shifted for a
  better comparison and the dashed black line marks the average
  energy, respectively.  Panel B: Radial distribution function between
  ammonia's nitrogen and water oxygen atoms in $NpT$ simulation using
  different potential models.}
\label{fig:pycharmm}
\end{figure}

\noindent
\cfont{Asparagus} also provides an interface between PhysNet and the
CHARMM simulation program via the pyCHARMM
API.\cite{Brooks.charmm:2009, pycharmm:2023} A trained ML-PES using
the PhysNet architecture predicts the required forces and atomic
charges, which are required for MD simulations and an electrostatic
interaction potential between the conformationally fluctuating NH$_3$
charges and static atomic point charges as defined by empirical force
fields such as CGenFF.\cite{mackerell2010CgenFF} Mechanically embedded
QM(ML)/MM simulations of ammonia (ML) in TIP3P water solvent (MM) were
performed. The necessary van-der-Waals parameters for the NH$_3$
molecule were those from
CGenFF.\cite{jorgensen1983tip3p,pycharmm:2023} Here, $NVT$ heating
simulation and $NPT$ equilibration simulations of a single ammonia
solute in 933 water solvent molecules was performed for 50\,ps each
followed by a 50\,ps $NVE$ simulation to check total energy
conservation (Figure \ref{fig:pycharmm}A) and simulation of and $NPT$
ensemble at 300\,K and normal pressure for 100\,ps with an MD-time
step of $\Delta t = 0.25$\,fs.  Listing \ref{code:pycharmm} shows an
extraction from the pyCHARMM script where an \cfont{Asparagus}
potential model is assigned.  It is to be mentioned that this
simulation setup of the strong base ammonia in water without allowing
proton transfer ($\ce{NH3 + H2O <=>> NH4+ + OH-}$) is chemically
inaccurate and only serves as a demonstration.\\

\noindent
Figure \ref{fig:pycharmm}A shows that the total energy is conserved
and only fluctuates with the same magnitude than in the $NVE$
simulations using the empirical CGenFF model (red line).  The total
energy sequence from simulations using the PhysNet model potential
(blue line) shows a slow oscillation around the energy average but
still within reasonable bounds. In addition, a
PaiNN\cite{schuett2021painn} ML-PES was trained on the metadynamics
data (B). Even though the forces RMSE of the PaiNN model with $24.5$
meV/\AA\/ is lower than the PhysNet model ($59.3$ meV/\AA), this is
not of concern here. Energy conservation is also observed in $NVE$
simulations using the PaiNN ML-PES (green line). Figure
\ref{fig:pycharmm}B shows the radial distribution functions ($g(r)$)
between the nitrogen atom of ammonia and the oxygen atoms of the water
solvent computed from the $NpT$ production simulation. The significant
difference is caused by the different atomic charge distribution of
the neutrally charged ammonia. Within the CGenFF model, the nitrogen
and hydrogen atoms of ammonia are assigned static point charges of
$-1.125$\,$e$ and $0.375$\,$e$, respectively. In comparison, the
trained PhysNet model predicts for ammonia in $C_{3v}$ equilibrium
geometry atomic charges of $-0.964$\,$e$ and $0.321$\,$e$, whereas the
trained PaiNN model predicts $-1.596$\,$e$ and $0.532$\,$e$.\\

\subsection{Chemical Reactions: Organometallic Complex}
To illustrate the capabilities of \cfont{Asparagus} to mix different
sampling methods for constructing a PES and its use in the study of a
reactive process, the hydrogen transfer step for the hydroformylation
catalytic cycle of alkenes using a simplified version of the Wilkinson
catalyst with cobalt (CoH(PMe$_{3})_{2}$(CO)) was
considered.\cite{wilkinson:1965} The training data for the reactive
step of interest was obtained by determining the reaction path between
intermediates taken from Ref. \citenum{van2022physics} with the NEB
method. For generating the training data, electronic structure
calculations were carried out at the PBE0/def2-SVP level including
D3BJ dispersion corrections using the ORCA code.\cite{orca:2020}
Subsequently, for each image along the NEB path, normal mode scanning
for all modes with frequencies larger than 100 cm$^{-1}$ was
performed. The total number of structures thus generated was 3069,
which was split into 80 \% for training, 10 \% for validation and 10
\% for testing. A PhysNet model was trained for 1000 epochs using a
batch size of 32. Using the obtained model, the minimum energy path
using NEB and the minimum dynamic path were determined. \\

\begin{figure}[h!]
\centering
\includegraphics[width=\textwidth]{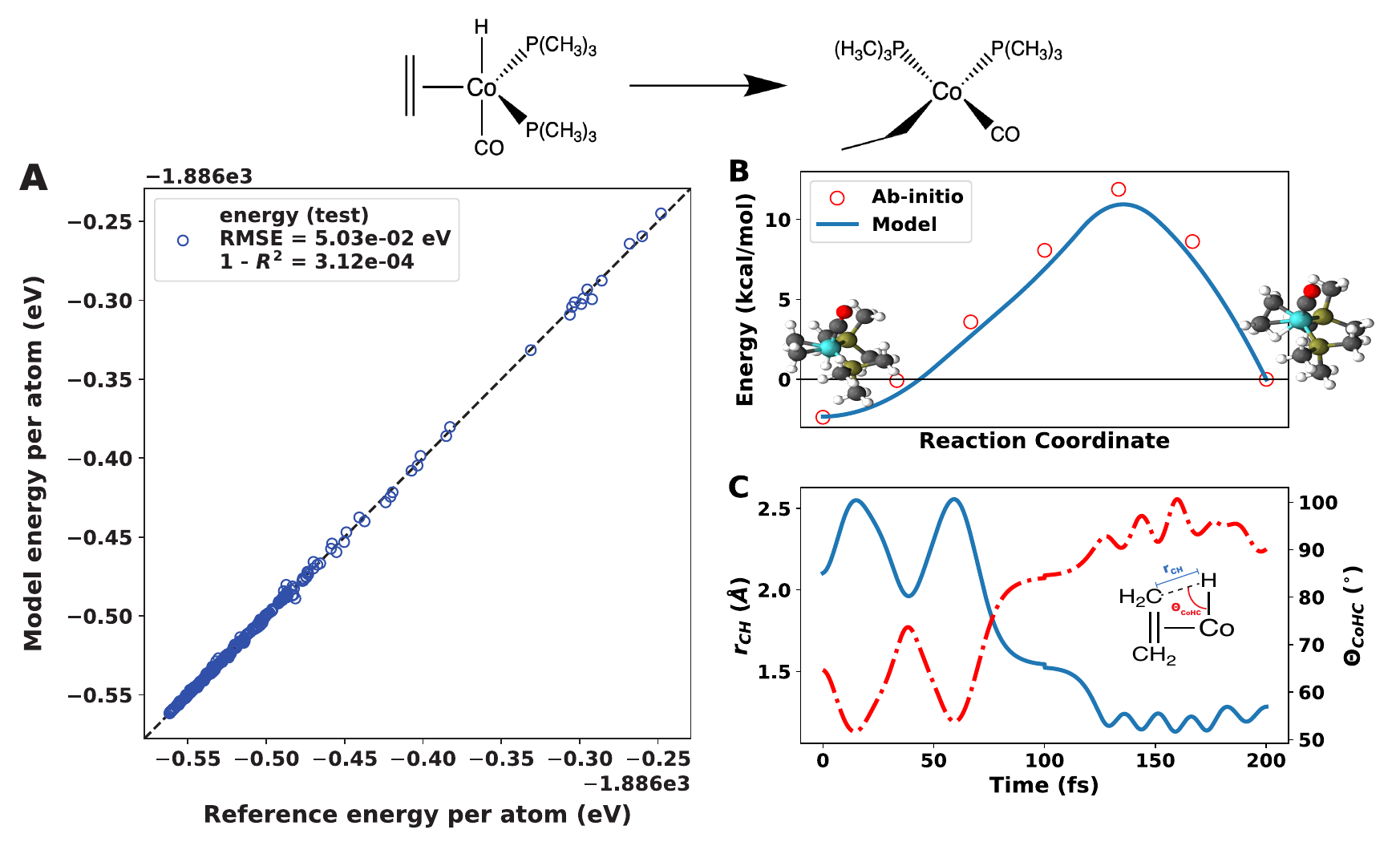}
\caption{Organometallic reaction. Panel A shows the correlation plot
  for the prediction energy in the test subset. Panel B displays the
  minimum energy path obtained from \textit{ab initio} calculations
  and, with the NN model, insight the panel structures of the
  equilibrium structures. Complementary panel C shows the change in
  the distance between the carbon in the alkene and the hydrogen atom
  bonded to the metal, as well as the angle between the C-H-Co atoms.}
\label{fig:organo}
\end{figure}

\noindent
The obtained ML-PES has RMSEs of 1.1 kcal/mol for the energies (Figure
\ref{fig:organo}A), 1.7 kcal/(mol \AA) for the forces, and 0.1 D for
the dipole moment in the test set. These results are near chemical
accuracy for energies and forces, indicating a good performance of the
fitted model. Further improvements of this ML-PES by increasing the
number of samples or using a different NN architecture are, of course,
possible. The layout of the graphics shown in Figure \ref{fig:organo}A
is automatically generated by \cfont{Asparagus} at the end of the test
procedure for the mentioned quantities. On request, \cfont{.csv} and
\cfont{.npz} files are also generated for each property, which
includes columns of reference and predicted values.\\

\noindent
Next, the MEP and MDP were determined using the tools implemented in
\cfont{Asparagus}. Figure \ref{fig:organo}B compares the MEP obtained
from \textit{ab initio} calculations (red circles) with that from the
fitted ML-PES (blue line). A good agreement between the two is found,
with a slight underestimation of the energies by the ML model. The
calculated MDP on the ML-PES, see Figure \ref{fig:organo}C, provides
chemical insight into the reactive process: as the distance between
the alkene C-atom and the H-atom attached to the Co-atom decreases, a
CH-bond (blue trace) is formed. Concomitantly, the C-H-Co angle
$\theta_{\rm CoHC}$ (red dot-dashed line) changes. Initially,
$\theta_{\rm CoHC} \sim 60^\circ$ because the hydrogen atom is closer
to the metallic centre. As the reaction progresses, the alkene rotates
to a position where the H bond is perpendicular to the Co and C
atoms. \\

\subsection{Handling Periodic Systems: Surfaces}
\label{ex3}
\cfont{Asparagus} also supports the training of ML-PESs for periodic
systems such as solids and surfaces. As an example, a PaiNN model is
trained to predict energies and forces of the diffusion motions of a
gold atom on an Al(100) surface. Reference data were obtained with the
GPAW\cite{mortensen2024gpaw} program package using the PBE density
functional and a projected augmented wave basis with an energy cutoff
at $200$\,eV. Energies and forces were computed with a $k$-points grid
of (4,4,1) with an Al(100) unit cell of size $2 \times 2 \times 3$ and
$4$\,\AA\/ vacuum level. As this is only an illustration, the
computational setup was neither optimized nor checked for
convergence. Rather, the computational setup was taken from the
surface diffusion tutorial reported by ASE\cite{ase:surfdiff} and
GPAW.\cite{gpaw:surfdiff} \\

\noindent
Reference data was obtained from NMS along the normal modes of the
gold atom and the four aluminum atoms of the top surface layer at
equilibrium conformations of gold at the hollow, bridge and top site
of the Al(100) surface. This yielded 5605, 6404 and 6254 reference
conformations, respectively. Further 208 conformations were generated
from metadynamics simulation whereby the CV was defined as the
separation between the gold atom and one of the top aluminum
atoms. The ML-PES was trained using PaiNN with a RMSE of $19$\,meV per
conformation for the test set (10\% of the overall 18471 reference
conformations).\\

\begin{figure}[h!]
\centering
\includegraphics[width=0.9\textwidth]{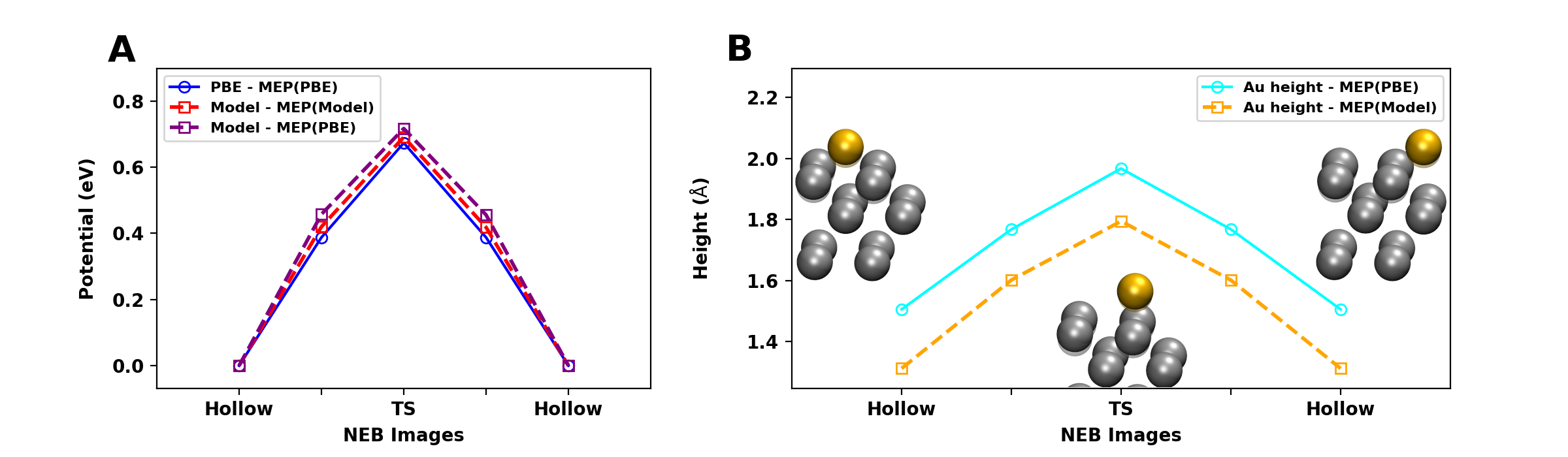}
\caption{Panel A: Minimum energy path (MEP) for a gold atom diffusion
  on Al(100) surface (hollow $\rightarrow$ bridge $\rightarrow$
  hollow) computed by NEB simulation using PBE/PAW level of theory
  (blue marker) and a PaiNN model potential (red marker) trained on
  Au/Al(100) reference data set. The purple line shows the energy
  prediction of the PaiNN model potential along the MEP structures
  from NEB with PBE/PAW.  Panel B shows the Au atom height from the
  ideal surface layer height and sketches of MEP.}
\label{fig:fig4}
\end{figure}

\noindent
Figure \ref{fig:fig4}A shows the diffusion potential of the gold atom
computed by nudged elastic band (NEB) method between two hollow
positions via the bridge position as transition state. The solid blue
line and dashed red line show the minimum energy path (MEP) computed
by the PBE/PAW reference method and trained model potential,
respectively.  With a reference diffusion barrier of $0.67$\,eV (solid
blue line at TS) the model potential predicts $0.71$\,eV for the same
conformation (dashed purple line at TS). The diffusion barrier in the
MEP obtained by the model potential is $0.69$\,eV (dashed red line at
TS).  Even though the diffusion barrier prediction in the model MEP is
close to the one of the reference MEP, the surface height of the gold
atom shown in Figure \ref{fig:fig4}B is predicted to be about
$0.17$\,\AA\/ lower in the model MEP than in the reference MEP.  The
reason is a qualitatively structural wrong equilibrium position of the
gold atom on the hollow site of Al(100). Such deficiency must be
corrected by, e.g., adaptive sampling of the PaiNN model potential
before it can be used for accurate predictions.  However, the example
shows the capability of \cfont{Asparagus} to handle periodic systems
including all functionalities also available for non-periodic
systems.\\

\section{Conclusions}
This work introduces the \cfont{Asparagus} workflow which supports a
largely autonomous construction of ML-PESs starting from a (few)
molecular structures.  The pipeline starts with different implemented
sampling methods, already existing datasets or other strategies to
obtain reference conformation for which, supported by
\cfont{Asparagus}, the reference properties can be computed with
\textit{ab-initio} codes.  Next, \cfont{Asparagus} handles the
generated data and makes them available for ML model training in a
\cfont{Asparagus} style database file.  \cfont{Asparagus} also shows
the statistical model performance metric in publication quality graphs
(c.f. Figure \ref{fig:organo}A).  At present, two popular atomistic NN
models, PhysNet and PaiNN, are available. Once the obtained ML-PES is
of the desired quality, \cfont{Asparagus} provides tools for
characterising the PES such as the calculation of MEP, MDP, DMC, or
harmonic frequencies. On the application-driven side,
\cfont{Asparagus} includes interfaces to ASE and pyCHARMM, allowing
the use of the generated potentials for running MD simulations. \\

\noindent
\cfont{Asparagus} provides a comprehensive workflow for autonomous
construction, validation, and use of ML-PESs. This considerably lowers
technical barriers increasing the confidence in model quality and
supporting workflow reproducibility, as well as the long-term
availability of the model generation pipeline. \cfont{Asparagus} is an
open project that will allow further improvements and incorporation of
the latest advances in the field of ML potentials models. Future
extensions include uncertainty
quantification,\cite{vazquez2022uncertainty} active learning,
automatization of transfer learning procedures, and interfaces to
other established MD codes such as LAMMPS.\cite{LAMMPS}

\section*{Acknowledgments}
The authors gratefully acknowledge financial support from the Swiss
National Science Foundation through grants $200020\_219779$,
$200021\_215088$, the NCCR-MUST, the AFOSR and the European Union's
Horizon 2020 research and innovation program under the Marie
Sk{\l}odowska-Curie grant agreement No 801459 -FP-RESOMUS. LIVS
acknowledges funding from the Swiss National Science Foundation (Grant
P500PN\_222297). The authors thank Dr. Silvan K\"aser for providing
the implementation regarding the computation of the Minimum Dynamic
Path and the Diffusion Monte Carlo method. The authors also
acknowledge Karthekan Sivasubramaniam and Timon Eya for the initial
tests of the code and members of the Meuwly group for helpful
discussions and testing.\\

\bibliographystyle{elsarticle-num}
\bibliography{refs}

\end{document}